\documentclass{iau}
\usepackage{graphicx}

\title[X-ray emission from hot accretion flows] 
{X-ray emission from hot accretion flows}

\author[Andrzej Nied\'zwiecki, Fu-Guo Xie \& Agnieszka St\c epnik]   
{Andrzej Nied\'zwiecki$^1$, Fu-Guo Xie$^2$ \& Agnieszka St\c epnik$^1$}
\affiliation{$^1$Department of Astrophysics, University of \L \'od\'z, Pomorska 149/153, 
90-236 \L \'od\'z, Poland\\ email: {\tt niedzwiecki@uni.lodz.pl} \\[\affilskip]
$^2$Key Laboratory for Research in Galaxies and Cosmology, Shanghai
  Astronomical Observatory, Chinese Academy of Sciences,
80 Nandan Road, Shanghai 200030, China\\ email: {\tt fgxie@shao.ac.cn}}

\pubyear{2014}
\volume{304}  
\pagerange{1--4}
\date{2013 December 12}
\jname{Multiwavelength AGN surveys and studies}
\editors{A. Mickaelian, F. Aharonian, D. Sanders, eds.}
\begin{document}

\maketitle

\begin{abstract}
Radiatively inefficient, hot accretion flows are widely considered as a relevant accretion mode in low-luminosity AGNs. We study spectral formation in such flows using a refined model with a fully general relativistic description of both the radiative (leptonic and hadronic) and hydrodynamic processes, as well as with an exact treatment of global Comptonization. We find that the X-ray spectral index--Eddington ratio anticorrelation as well as the cut-off energy measured in the best-studied objects favor accretion flows with rather strong magnetic field and with a weak direct heating of electrons. Furthermore, they require a much stronger source of seed photons than considered in previous studies. The nonthermal synchrotron radiation of relativistic electrons seems to be the most likely process capable of providing a sufficient flux of seed photons. 
Hadronic processes, which should occur due to basic properties of hot flows, provide an attractive explanation for the origin of such electrons.
\end{abstract}


\noindent
{\underline{\it The hot flow model}}

The most developed and commonly accepted model of low-luminosity black hole systems is that of optically thin, two-temperature accretion flows (ususally referred to as ADAFs), see, e.g., \cite[Narayan \& McClintock (2008)]{nmc08} for a review. Thermal Comptonization in such flows is then considered as the origin of strong X-ray emission, often dominating the radiative output of low-luminosity systems. The major problem for the model, pointed out e.g.\ by \cite[Yuan \& Zdziarski (2004)]{yz04}, is that electron temperatures predicted by the model are much higher than  temperatures indicated by the high-energy cut-offs in the observed spectra. However, previous comparisons with the data relied mostly on simplified models, in particular involving the use of a pseudo-Newtonian potential  as well as local approximations of Comptonization, both being potential sources of serious inaccuracies.

In a standard ADAF theory, seed photons are provided by thermal synchrotron radiation. The weak efficiency of this 
emission process leads to disagreement with observations (discussed below). Irradiation of the hot flow by thermal emission from a surrounding, cold disc may provide an additional strong flux of soft photons and this process is possibly responsible for softening observed above $\sim 1$\% of the Eddington luminosity, $L_{\rm Edd}$   (illustrated in Fig.\ 2a); at lower luminosities this is, however, a negligible effect. A rather weakly explored solution for the seed photons problem involves nonthermal synchrotron emission of relativistic electrons. Such electrons may be present due to either nonthermal acceleration processes or $\pi^\pm$ decay. The latter is a natural process in ADAFs, and we focus on it here, as the two-temperature structure is their essential feature and proton-proton interactions should lead to a substantial production of pions.

\begin{figure}[b]
\begin{center}
\includegraphics[height=4.4cm]{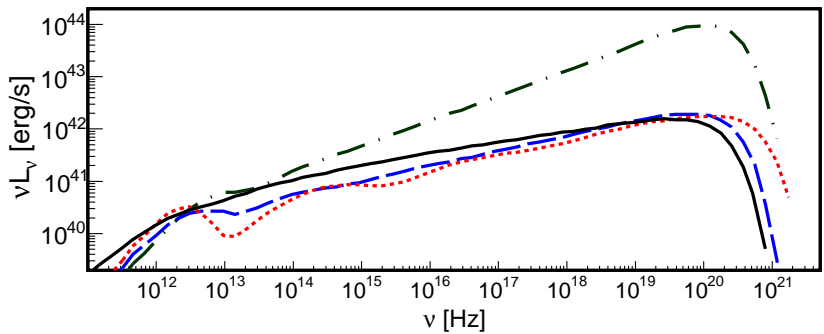}
\caption{
Spectra of radiation produced in a hot flow with $M =2 \times 10^8 {\rm M}_{\odot}$, $\dot m=0.1$ and $a=0.95$. The solid spectrum is for ($\beta,\delta$)=($1,10^{-3}$) in the model with hadronic $e^\pm$. The remaining spectra are for ($\beta,\delta$) = ($1,10^{-3}$;   dashed), ($9,10^{-3}$;  dotted) and ($1,0.5$; dot-dashed) in the purely thermal model. 
}
\label{fig1}
\end{center}
\end{figure}

We have recently developed the model with an exact treatment of relevant radiative processes, see \cite[Xie et al.\ (2010)]{x10}, \cite[Nied\'zwiecki et al.\ (2012)]{n12}, \cite[Nied\'zwiecki et al.\ (2013)]{n13}. Below we briefly describe the dependence on main parameters (illustrated in Fig.\,\ref{fig1}), then, we compare  the model predictions with observations. 
The standard version of the model, for which the thermal synchrotron radiation is the only source of seed photons, is referred to as a {\it purely thermal model}. The extension of the model, taking into account additional seed photons from nonthermal synchrotron of relativistic $e^\pm$ produced by $\pi^\pm$  decay, is referred to as a {\it model with hadronic $e^\pm$}. 

We use two parameters to describe poorly understood plasma physics:
the fraction of the dissipated energy which heats directly electrons is denoted by $\delta$ and the magnetic field strength is parametrized by the ratio of gas to magnetic pressure, $\beta$. Both $\beta$ and $\delta$ appear crucial for spectral properties. 
In models with large $\delta$ ($\simeq 0.5$), luminosity is  larger by an order of magnitude, and electron temperature is a few times higher, than in models with small $\delta$ ($\simeq 10^{-3}$). For $\delta=0.5$, the temperature is $\sim 500$ keV or higher regardless of other parameters. At small $\delta$,
$\beta$ is the key parameter for the plasma temperature. Apart from the obvious dependence of the synchrotron emissivity on the magnetic field strength, an even stronger effect results from a higher  compressibility of  flows with smaller $\beta$, making such flows closer to a slab geometry; in turn, flows with larger $\beta$ are closer to a spherical geometry, for which Compton cooling is much less efficient.

Our model allows to study the dependence on the black hole spin, parametrized here by
$a = J / (c R_{\rm g} M)$, where $M$ and $J$ are the  black hole mass and angular momentum and $R_{\rm g}=GM/c^2$. In general it is an important parameter, in particular determining the radiative efficiency of flows with large $\delta$ as well as affecting the value of the critical accretion rate, above which a hot flow ceases to exist. On the other hand, the value of $a$ does not affect our constraints on $\delta$ and $\beta$.

Large spectral changes are supposed to result from changes of the accretion rate $\dot{M}$, given below 
by $\dot m = \dot M / \dot{M}_{\rm Edd}$, where $\dot{M}_{\rm Edd}= L_{\rm Edd}/c^2$.
Increase of $\dot m$ yields a stronger heating and, hence, a larger luminosity and a harder X-ray spectrum. Despite a stronger heating, however, the electron temperature decreases with increasing $\dot m$ due to the increase of the optical depth, $\tau$, which strongly enhances the Compton cooling.

Finally, we find that the synchrotron emission of pion-decay electrons reduces temperature for small $\delta$  and rather small $\beta$ ($\simeq 1$); for larger $\beta$ or $\delta$, differences between the two versions of the model, purely thermal and with hadronic $e^\pm$, are insignificant  (in such cases the nonthermal synchrotron is comparable to the  thermal synchrotron emission, which is  strong at $T \sim 500$ keV).
The rate of pion production, which determines the seed photon flux in model with hadronic $e^\pm$, is proportional to the square of density, so the difference between the two versions of the model increases with increasing $\dot m$.

\begin{figure}[b]
\begin{center}
 \includegraphics[height=7.4cm]{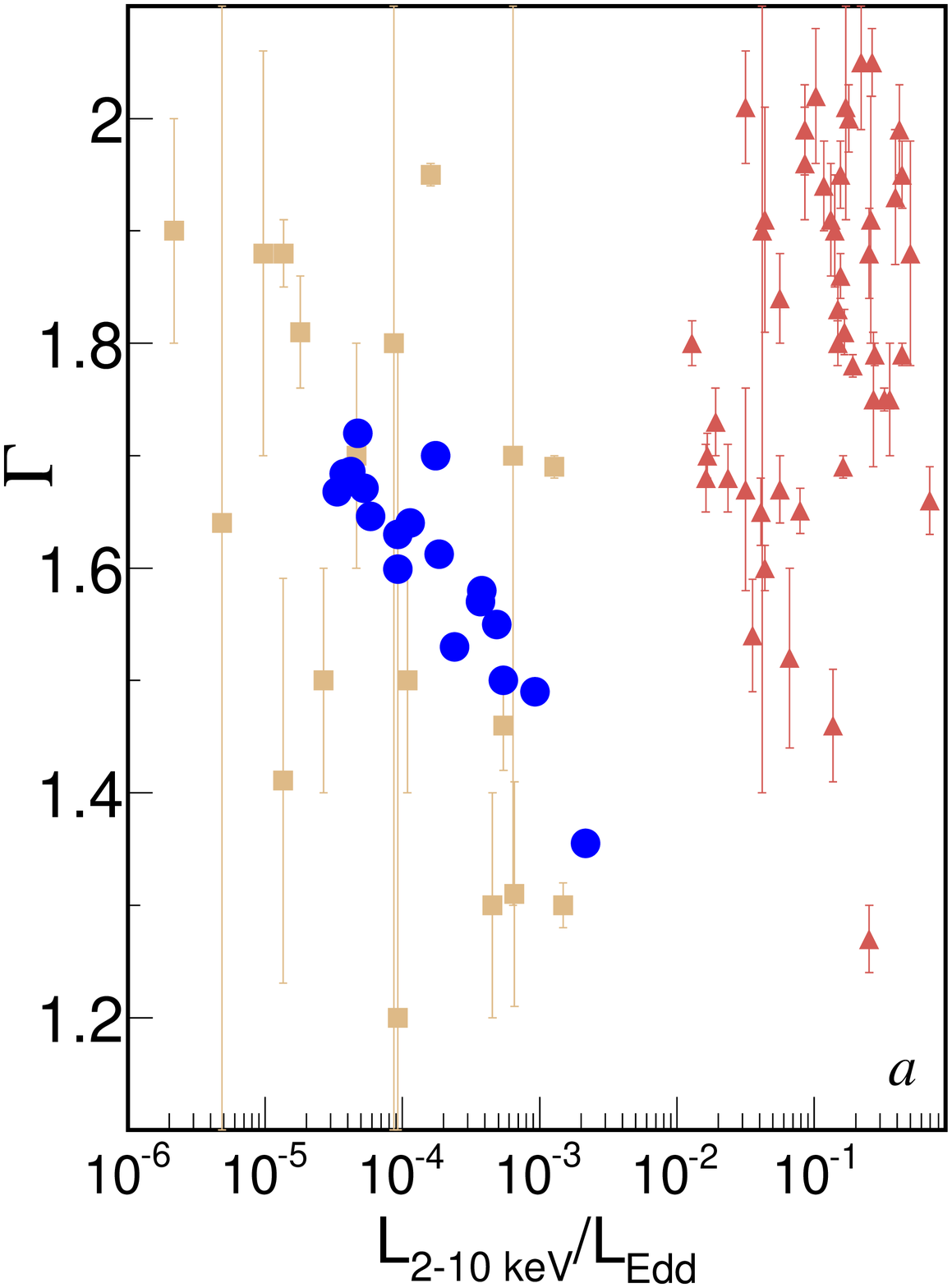}\includegraphics[height=7.4cm]{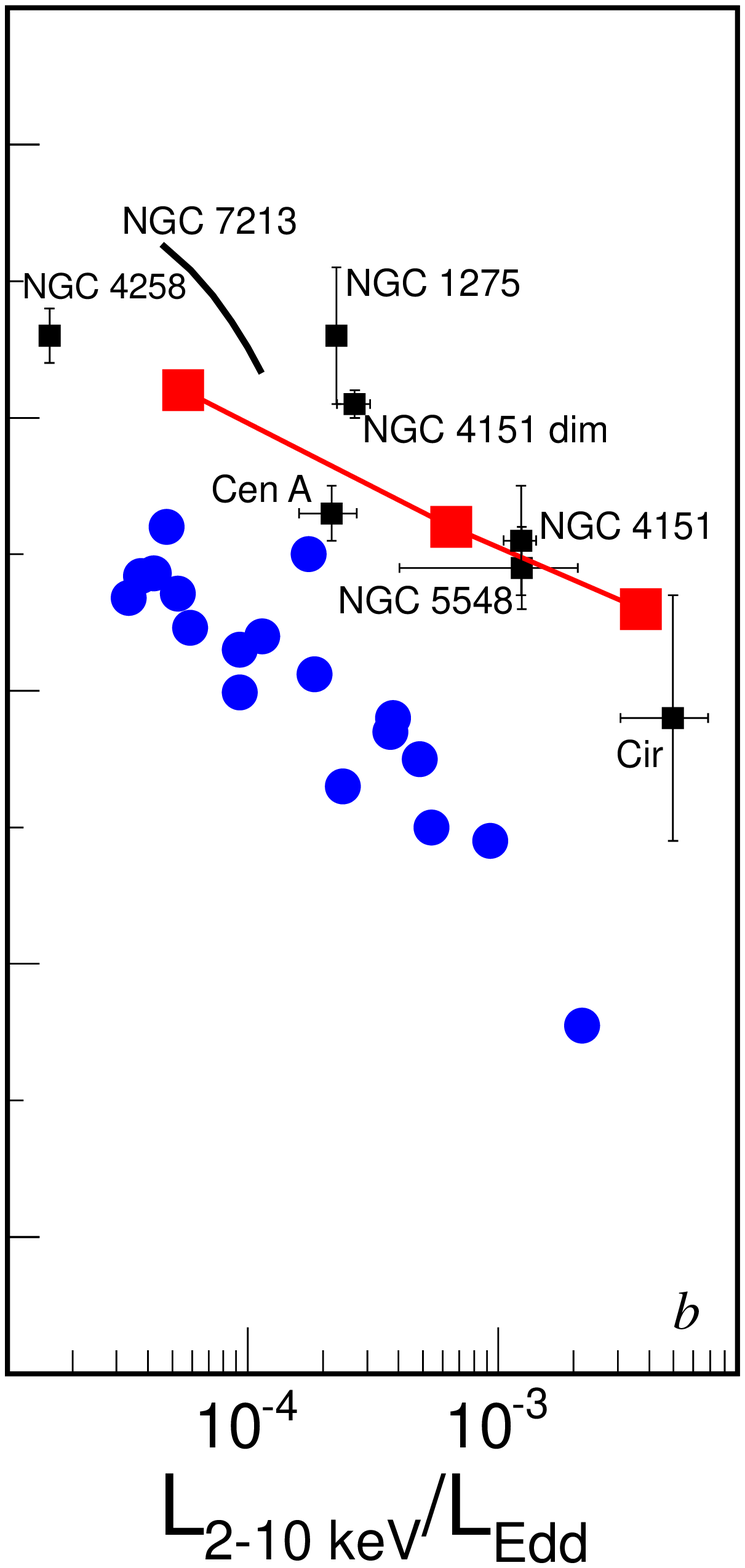}\hspace{0.05cm}\includegraphics[height=7.3cm]{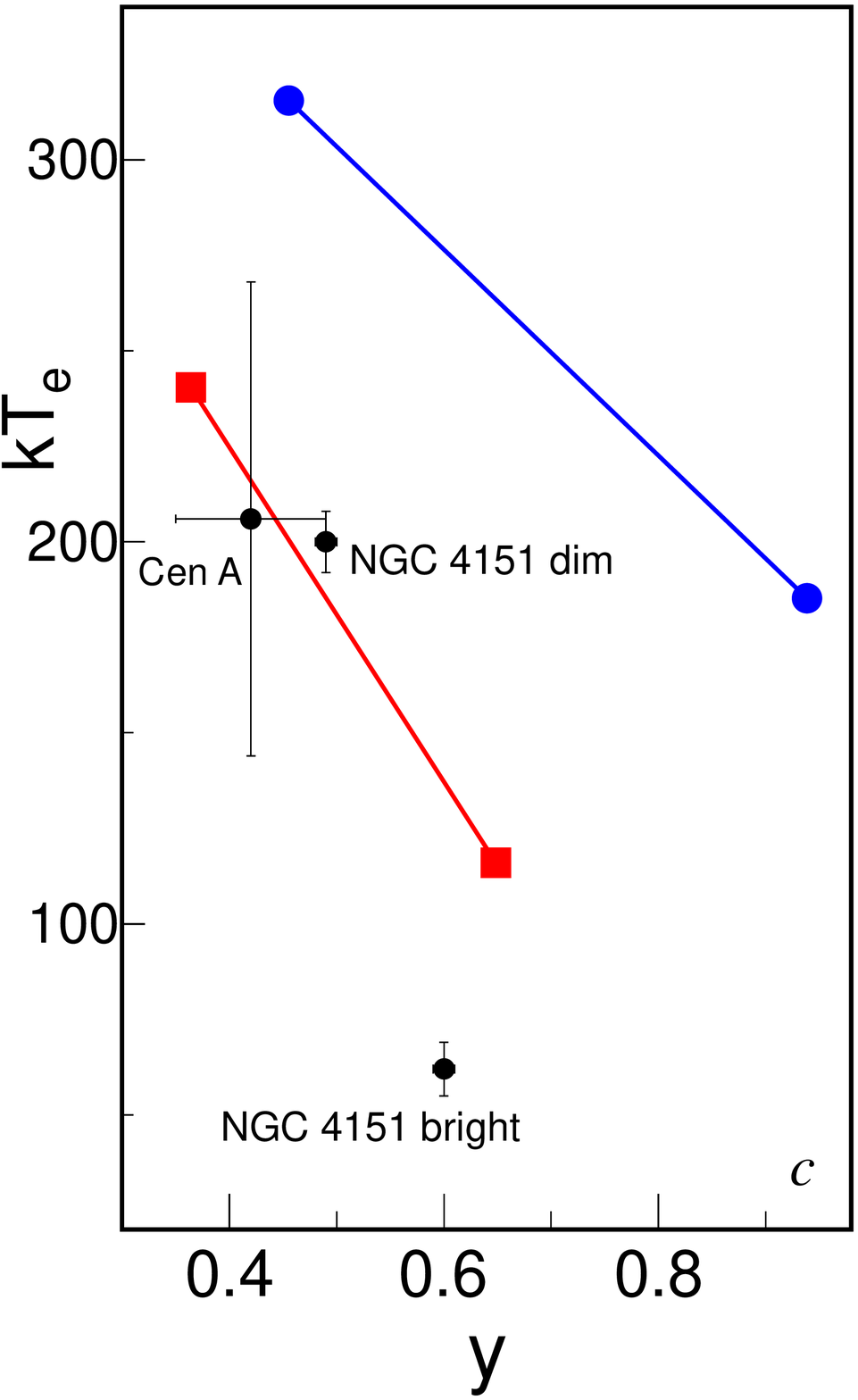}
 \caption{Observational data for low-luminosity AGNs compared with predictions of the hot flow model. (a,b) show the photon spectral index as a function of the 2--10 keV Eddington ratio. Squares in (a) shows the data for low-luminosity Seyferts from \cite[Gu \& Cao (2009)]{gc09} and triangles show the data for luminous Seyferts from \cite[Zhou \& Zhang (2010)]{zz10}. Small squares in (b) show the data for our high-quality sample (see text). Circles in (a) and (b) show the purely-thermal model points for broad ranges of $\beta$, $\delta$, $\dot m$ and $a$ (see text). Big squares, connected by the solid line in (b),  are for $\beta=1$, $\delta=10^{-3}$, $\dot m=0.1$, 0.3 and 0.45, from left to right, in the model with hadronic $e^\pm$. (c) shows electron temperatures and Compton parameters for two best-studied AGNs; circles and squares show the model points, purely-thermal and with hadronic $e^\pm$, respectively,  for $\beta=1$, $\delta=10^{-3}$, $\dot m=0.1$ (left) and 0.3 (right).}
   \label{fig2}
\end{center}
\end{figure}

\noindent
{\underline{\it Comparison with observations}}

For the comparison we use the photon spectral index in the X-ray range, $\Gamma$, the Compton parameter, $y$ ($\equiv 4 \tau kT_{\rm e}/m_{\rm e}c^2$) and the Eddington ratio, $\lambda = L_{2-10}/ L_{\rm Edd}$, where $L_{2-10}$ is the 2--10 keV luminosity of the flow. The $\Gamma$--$\lambda$ anticorrelation was revealed in recent years in several studies of low-luminosity  (with $\lambda < 0.01$) systems. Such a negative relation is seen both in individual AGNs (e.g., \cite[Lubi\'nski et al.\,(2010)]{l10}) and in large samples of AGNs (e.g.\ \cite[Gu \& Cao (2009)]{gc09}). The latter show also that above $\lambda \sim 10^{-3}$--$10^{-2}$ the relation is inverted and the X-ray spectra soften with increasing luminosity. 
In Fig.\,\ref{fig2}a we show the data for low-luminosity Seyfert galaxies from \cite[Gu \& Cao (2009)]{gc09} compared with predictions of our model; the data for luminous Seyfert galaxies from \cite[Zhou \& Zhang (2010)]{zz10} are also plotted to illustrate the invertion of the $\Gamma$--$\lambda$ relation.
Studies of large samples of AGNs, such as these adopted by \cite[Gu \& Cao (2009)]{gc09}, usually use a simplified approach to spectral modeling,  which allows to estimate generic trends, however, the assumed models are too simple to be used  for precise comparisons with physical models.
Then, we gathered  the values of the intrinsic X-ray slopes from detailed models of several well studied objects (called below a high-quality sample): 
NGC 4151 (\cite[Lubi\'nski et al.\,(2010)]{l10}), NGC 5548 (\cite[Brenneman et al.\,(2012)]{b12}), NGC 4258 (\cite[Yamada et al.\,(2009)]{y09}), Circinus (\cite[Yang et al.\,(2009)]{yang09}), Centaurus A (\cite[Beckmann et al.\,(2011)]{b11}), NGC 1275 (\cite[Donato et al.\,(2004)]{d04}) and NGC 7213 (fit to $\Gamma(\lambda)$; \cite[Emmanoulopoulos et al.\,(2012)]{e12}). These data points are shown in Fig.\,\ref{fig2}b.
We see that the high-quality sample shows a much tighter correlation between $\Gamma$ and $\lambda$ than that apparent from  Fig.\,\ref{fig2}a.

Circles in Figs \ref{fig2}ab show the model points for the purely-thermal model with $0.03\le \dot m \le 0.3$, $0\le a  \le 0.998$, $1 \le \beta \le 9$ and $10^{-3} \le \delta \le 0.5$. As we can see, for such a broad range of parameters the model gives a rather narrow distribution of $\Gamma$ at fixed $\lambda$. The comparison in Fig.\,\ref{fig2}a is not particularly conclusive. In Fig.\,\ref{fig2}b (which we regard to be more reliable) we see that the purely-thermal  model predicts spectra systematically harder than those of the high-quality sample. As also seen in Fig.\,\ref{fig2}b,  the hot flow model can be reconciled with the data if the pion-decay electrons are taken into account. We emphasize again that small $\delta$ and small $\beta$ are crucial for achieving this agreement.

The above finding is further confirmed by comparing the model predictions with electron temperatures measured in AGNs. Here the observational grounds are even more uncertain than for the intrinsic X-ray slope; accurate assessment of the plasma temperature requires high-quality data around the high-energy cut-off, which are available only for a few objects. 
The most precise measurements were done for NGC 4151 (\cite[Lubi\'nski et al.\,(2010)]{l10}) and Cen A (\cite[Beckmann et al.\,(2011)]{b11}), the results are shown in Fig.\,\ref{fig2}c. 
We see that the purely-thermal model with $\beta=1$ and $\delta=10^{-3}$ overestimates the temperature and we note that all models with large $\delta$ as well as these with large $\beta$  predict $kT > 320$ keV, i.e.\, above the upper boundary of the figure. 
The model with hadronic $e^\pm$ is in good agreement  with the data at $\lambda \sim 10^{-4}$; at $\lambda \sim 10^{-3}$ the model temperature is still larger than that measured in the bright state of NGC 4151. This may indicate that a stronger magnetic field, with $\beta < 1$, is required at this $\lambda$.

Interestingly, Cen A and NGC 1275, which are FR I radiogalaxies, seem to follow a similar $\Gamma(\lambda)$ relation as seen in the best-studied Seyferts in this range of $\lambda$. This supports the accretion flow (rather than jet) origin of the X-ray emission of FR Is at such $\lambda$, indicated also by the thermal-like cut-off in Cen A (\cite[Beckmann et al.\,(2011)]{b11}). Also, consistently with our requirement of small $\delta$, radiative efficiencies assessed for FR Is are small (e.g.\ \cite[Donato et al.\,(2004)]{d04}).
\begin{acknowledgements}

This research has been supported by the Polish NCN grant N N203 582240.
FG Xie is supported by the Natural Science Foundation of China (grants 11103061 \& 11203057).
\end{acknowledgements}


\begin{thebibliography}{14}

\bibitem[Beckmann et al. (2011)]{b11}
{Beckmann V., et al.} 2011 \textit{AA}, 531, 70

\bibitem[Brenneman et al.\,(2012)]{b12}
{Brenneman L., et al.} 2012,  \textit{ApJ}, 744, 13

\bibitem[Donato et al. (2004)]{d04}
{Donato D., Sambruna R.~M.,  \& Gliozzi M.} 2004 \textit{ApJ}, 617, 915


\bibitem[Emmanoulopoulos et al. (2012)]{e12}
{Emmanoulopoulos D., et al.} 2012, \textit{MNRAS}, 424, 1327


\bibitem[Gu \& Cao (2009)]{gc09}
{Gu M.\,\& Cao X.}  2009, \textit{MNRAS}, 399, 349 


\bibitem[Lubi\'nski et al.\,(2010)]{l10}
{Lubi\'nski P., et al.} 2010, \textit{MNRAS}, 408, 1851


\bibitem[Narayan \& McClintock (2008)]{nmc08}
{Narayan, R., \& McClintock, J.E.} 2004, \textit{New Astron. Revs}, 51, 733


\bibitem[Nied\'zwiecki et al (2013)]{n13}
{Nied\'zwiecki A., Xie F.-G., \& St\c epnik A.} 2013, \textit{MNRAS}, 432, 1576

\bibitem[Nied\'zwiecki et al (2012)]{n12}
{Nied\'zwiecki A., Xie F.G., \& Zdziarski A.} 2012, \textit{MNRAS}, 420, 1195


\bibitem[Xie et al.\,(2010)]{x10}
{Xie F.-G., et al.} 2010, \textit{MNRAS}, 403, 170

\bibitem[Yamada et al. (2009)]{y09}
{Yamada S., et al.} 2009, \textit{PASJ}, 61, 309

\bibitem[Yang et al. (2009)]{yang09}
{Yang et al.} 2009, \textit{ApJ}, 691, 131


\bibitem[Yuan \& Zdziarski (2004)]{y04}
{Yuan F.\,\& Zdziarski A.} 2004, \textit{MNRAS}, 354, 953

\bibitem[Zhou \& Zhang (2010)]{zz10}
{Zhou X.-L.\,\& Zhang S.-N.} 2010, \textit{ApJ}(Letters), 713, L11

\end{thebibliography}
\end{document}